\documentclass[review]{elsarticle}

\usepackage{lineno,hyperref}
\usepackage{epstopdf}
\modulolinenumbers[5]

\journal{Journal of \LaTeX\ Templates}

\begin{document}

\begin{frontmatter}

\title{Charge Carrier Dynamics of the Heavy-Fermion Metal CeCoIn$_5$ Probed by THz Spectroscopy}

\author{Uwe S. Pracht}\ead{uwe.pracht@pi1.physik.uni-stuttgart.de}
\author{Julian Simmendinger, Martin Dressel}
\address{1. Physikalisches Institut, University of Stuttgart, D-70569 Stuttgart, Germany}
\author{Ryota Endo, Tatsuya Watashige, Yousuke Hanaoka, Masaaki Shimozawa}
\address{Department of Physics, Kyoto University, Kyoto 606-8502, Japan}
\author{Takahito Terashima}
\address{Research Center for Low Temperature and Materials Science, Kyoto University, Kyoto 606-8502, Japan}
\author{Takasada Shibauchi}
\address{Department of Advanced Materials Science, University of Tokyo, Chiba 277-8561, Japan}
\author{Yuji Matsuda}
\address{Department of Physics, Kyoto University, Kyoto 606-8502, Japan}
\author{Marc Scheffler}\ead{marc.scheffler@pi1.physik.uni-stuttgart.de}
\address{1. Physikalisches Institut, University of Stuttgart, D-70569 Stuttgart, Germany}

\begin{abstract}
We discuss the charge carrier dynamics of the heavy-fermion compound CeCoIn$_5$ in the metallic regime measured by means of quasi-optical THz spectroscopy. The transmittance of electromagnetic radiation through a CeCoIn$_5$ thin film on a dielectric substrate is analyzed in the single-particle Drude framework. We discuss the temperature dependence of the electronic properties, such as the scattering time and dc-conductivity and compare with transport measurements of the sheet resistance. Towards low temperatures, we find an increasing mismatch between the results from transport and Drude-analyzed optical measurements and a growing incapability of the simple single-particle picture describing the charge dynamics, likely caused by the evolving heavy-fermion nature of the correlated electron system.   
\end{abstract}

\end{frontmatter}

\section{Introduction}
\label{i}
Amongst the different heavy-fermion materials, CeCoIn$_5$ has a prominent role as the Ce-based heavy-fermion superconductor with the highest $T_c$~= 2.3~K and as the most studied material of the \lq 115\rq -family of Ce compounds.\cite{petrovic2001,sarrao2007,thompson2013} Detailed information about electronic excitations and charge dynamics in heavy fermions can be obtained by optical spectroscopy in different spectral ranges.\cite{basov2011,scheffler2013} Concerning infrared optics, CeCoIn$_5$ has been examined in great detail.\cite{singley2002,mena2005,burch2007,okamura2015} These studies focused on signatures of the hybridization of conduction and $f$ electrons whereas the dynamics of the mobile charge carriers in heavy-fermion materials have to be probed at even lower frequencies, in the GHz and THz ranges.\cite{scheffler2013,webb1986,degiorgi1999,marc_nature,dressel2006,scheffler2006,scheffler10} Previous GHz and THz studies on CeCoIn$_5$ have explicitly addressed the superconducting transition,\cite{ormeno2002,nevirkovets2008,SudhakarRao2009,truncik2013} while the heavy-electron charge dynamics in the metallic state of CeCoIn$_5$ have hardly been addressed by optics so far. The main reason here is that the sensitivity of broadband GHz and THz experiments is limited and for studies on highly conductive materials requires thin-film samples,\cite{marc_rsi,marc_strip,Pra13}  which are difficult to grow in the case of heavy-fermion metals. Here we explicitly study the charge response of a CeCoIn$_5$ thin film.

\section{Methods}
\label{m}
The sample under study is a 70\,nm thick film of CeCoIn$_5$ deposited via molecular beam epitaxy on a dielectric $5\times5\times0.5\,$\,mm$^3$ MgF$_2$ substrate. This technique has already been applied for growth of several thin-film systems based on CeCoIn$_5$ and CeIn$_3$.\cite{shishido2010,mizukami2011,shimozawa2012,goh2012,shimozawa2014}
Other than in previous THz studies \cite{Sch13} on CeCoIn$_5$ thin films, we do not need additional metallic buffer layers to achieve high-quality samples. THz transmission measurements, however, still remain challenging for several reasons. First, the film needs to be thin enough to allow for a detectable transmission signal, while the sample quality favors thick films. Here we have chosen a thickness of 70\,nm, which is a compromise between sample quality and suitability for our experimental technique. Second, thin films of CeCoIn$_5$ often rapidly degrade in ambient air conditions so that exposure time must be cut to a minimum.\cite{Sch13} Third, the aperture, through which the focused THz radiation passes before it is transmitted through the sample, needs to have a diameter $d_a$ smaller than the sample. We have chosen $d_a=$ 3\,mm which restricts our accessible spectral range to wavelengths shorter than 1\,mm due to diffraction effects.

 After deposition in Kyoto, the film was sealed in a glass tube under vacuum conditions before it was shipped to Stuttgart. Right after removal from the glass tube, the sample was mounted onto the THz sample holder, transferred to the cryostat, and rapidly cooled down in He-gas atmosphere. Here, the overall exposure time to ambient air was less than 5 minutes. The entire optical measurements were subsequently performed during a period of about 36 hours. During this time, the sample was always kept below 150\,K. Afterwards, it was removed from the cryostat and contacted in standard 4-point geometry in order to measure the dc-sheet resistance, and correspondingly obtain the dc transport resistivity $\rho_{dc}$. Even after a measurement time of $\sim$\,60 hours including an exposure time of $\sim$\,30 minutes we could not infer any signs of degradation from the THz spectra.

\begin{figure}
\begin{centering}
\includegraphics[scale=0.3]{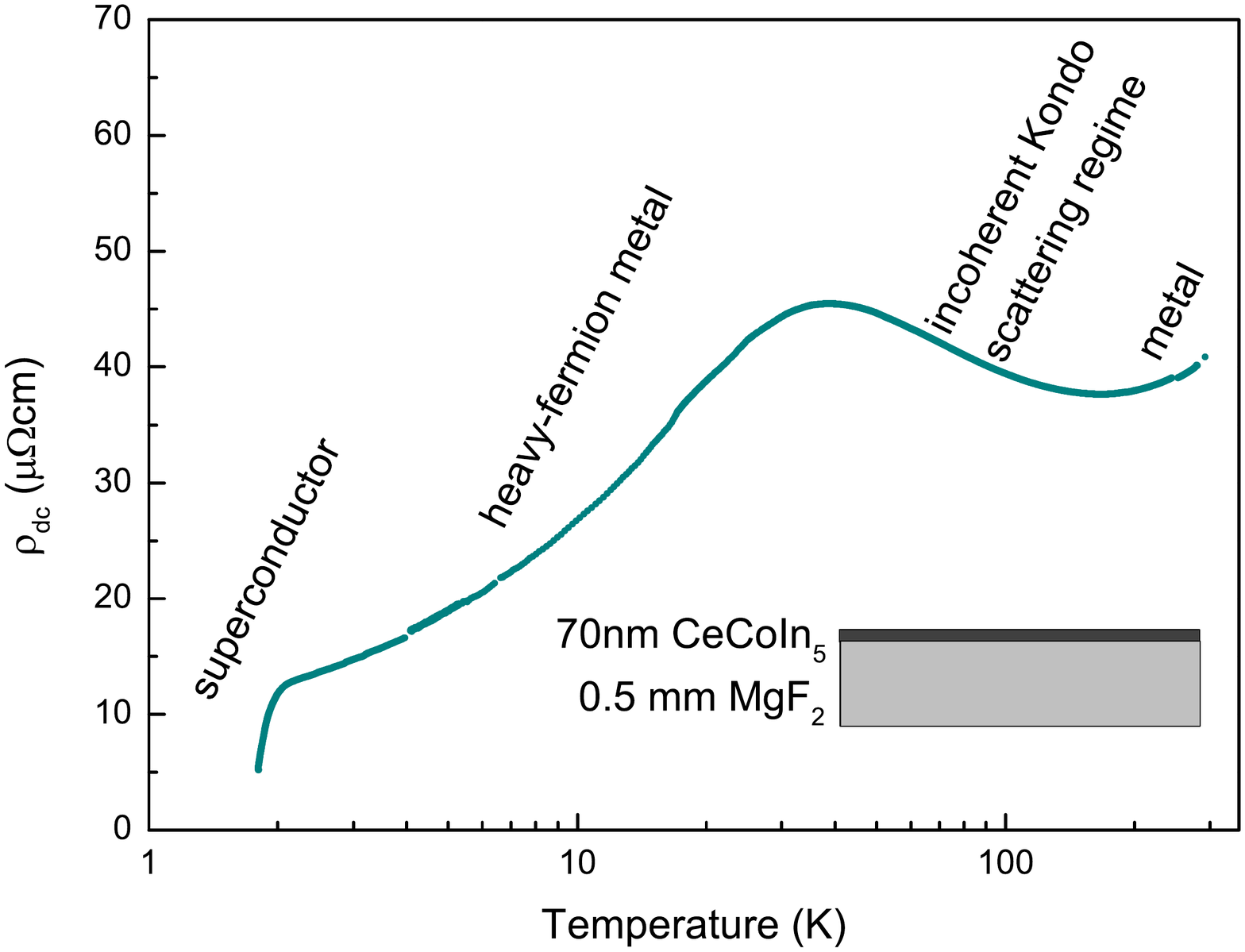}
\caption{\label{res} (Color online) dc transport resistivity $\rho_{dc}$ versus temperature of the CeCoIn$_5$ film studied in this work. Though being rather thin, all characteristics known from single-crystal CeCoIn$_5$ are well recovered. The inset shows a schematic drawing of the bilayer system.  }
\end{centering}
\end{figure}

\begin{figure}
\begin{centering}
\includegraphics[scale=0.15]{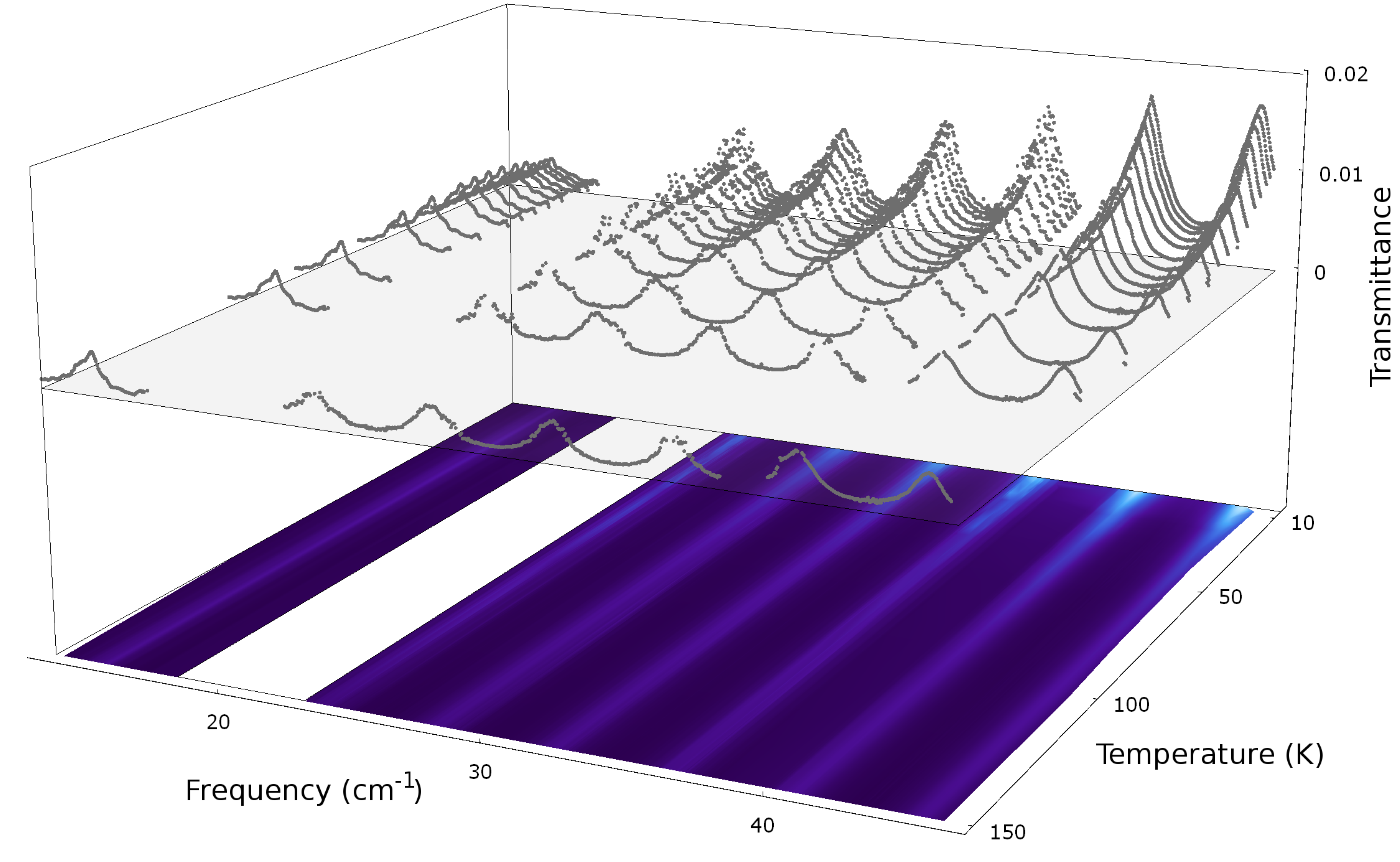}
\caption{\label{Tr_SO34} (Color online) Raw transmittance of 70\,nm CeCoIn$_5$ film as function of temperature and frequency. The pronounced oscillation pattern is caused by the dielectric substrate acting as a Fabry-Perot resonator. At high temperatures, the oscillation pattern is constant while it acquires a strong frequency dependence towards low temperatures: the transmittance increases with increasing frequency and decreasing temperature except for the low frequency- and temperature limit, where it is suppressed.}
\end{centering}
\end{figure}

The transmittance was measured using a set of tunable backward-wave oscillators as sources of coherent and monochromatic THz radiation and a He-cooled bolometer as detector.\cite{Pra13} Measurements were conducted in a spectral range spanning 13 - 46\,cm$^{-1}$ (i.e. 0.4 - 1.4\,THz). Sample temperatures between 6\,K and 150\,K were maintained in a home-built cryostat. We omitted measurements at higher and lower temperatures in order to restrain the measurement time and avoid degradation. Since our transmittance data is for a two-layer system (substrate + film), we measured a bare reference substrate in the same run to disentangle the properties of both layers.  
	%%%%%%%%%%%%%%%%%%%%%%%%%%%%%%
	%%%%%%%%%%%%%%%%%%%%%%%%%%%%%%
	%%%%%%%%%%%%%%%%%%%%%%%%%%%%%%
	%%%%%%%%%%%%%%%%%%%%%%%%%%%%%%
	%%%%%%%%%%%%%%%%%%%%%%%%%%%%%%
	%%%%%%%%%%%%%%%%%%%%%%%%%%%%%%

\section{Results and Discussion}
\label{r}
The temperature dependence of the dc transport resistivity $\rho_{dc}$ is shown in Fig. \ref{res}. The sample exhibits all characteristic regimes well known for CeCoIn$_5$,\cite{petrovic2001,malinowski2005} which is consistent with an excellent film quality. Starting at room temperature, the system behaves like a normal metal and $\rho_{dc}$ decreases slightly, passes a minimum at around $T_{min}=165$\,K and then increases again due to incoherent Kondo scattering. This increase levels off at around $T_{max}=40$\,K, where the system enters the coherent heavy-fermion state which then goes along with a rapid reduction of $\rho_{dc}$ upon further cooling before the curve bends down to the superconducting transition at presumably $T_c\approx 1.8$\,K slightly below our lowest measured temperature.

\begin{figure}[tb]
\begin{centering}
\includegraphics[scale=0.45]{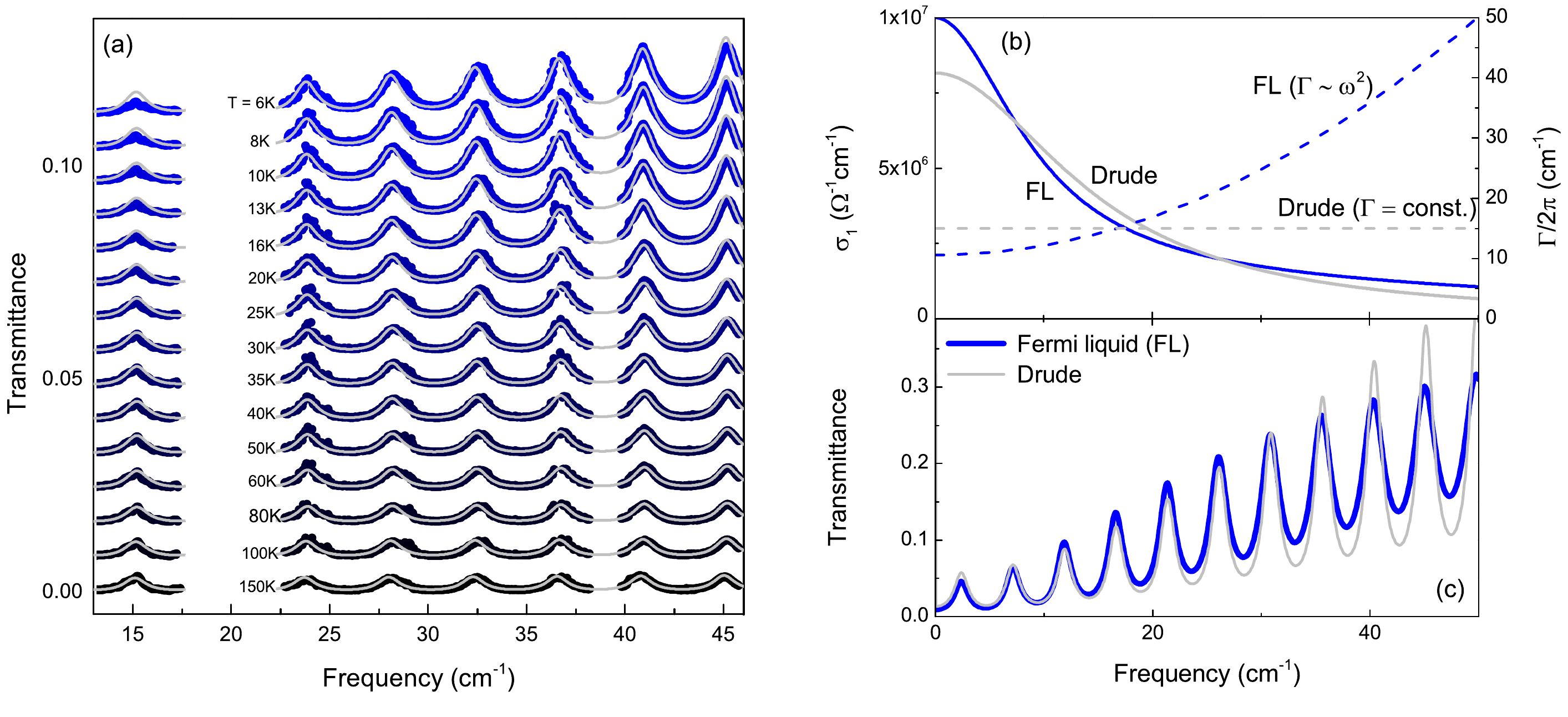}
\caption{\label{fits} (Color online) (a) Transmittance spectra and Fresnel fits including Drude conductivity for several temperatures. The spectra and fits are shifted for clarity. While at high temperatures the transmittance is well described by the fit, the fidelity gradually decreases towards low temperatures signaling a breakdown of the simple single-particle Drude description. (b) and (c) schematically explain such behavior by comparing the Drude and Fermi-liquid (FL) predictions for the optical response: in FL theory the scattering rate $\Gamma$ (dashed) is frequency dependent, which leads to different spectra in the real part $\sigma_1$ of the conductivity (full) and in the transmittance.}
\end{centering}
\end{figure}

The raw transmittance is displayed in Fig. \ref{Tr_SO34} as function of frequency and temperature. The spectra feature pronounced Fabry-Perot (FP) oscillations that stem from multiple reflections inside the substrate.\cite{Pra13} At $T=150$\,K, the highly-conductive metallic film suppresses the overall transmittance (at the FP peaks) to less than 1\%. Upon cooling, the spectra acquire a strong frequency dependence beyond the FP pattern. This is most pronounced in the high-frequency limit, where at $T=6$\,K the transmittance is about 4 times larger than at high temperatures. At intermediate frequencies, this enhancement is less strong and at the lowest frequencies it is even reversed at around 30\,K. Measurements of the bare substrate reveal only minor losses at 150\,K, which disappear below $\sim$80\,K. Thus, the observed frequency and temperature dependence of the transmittance can completely be attributed to the electronic properties of the CeCoIn$_5$ film. Such a behavior at THz frequencies is expected for a good metal, where the electron scattering rate $\Gamma = 1/\tau$, with $\tau$ the time between two scattering events, shifts into the examined spectral range upon cooling. In our case of CeCoIn$_5$ the shift of $\Gamma$ is attributed to the gradual emergence of a coherent heavy-fermion state with a concomitant slowing down of the Drude relaxation rate combined with a reduction of temperature-dependent scattering e.g.\ due to phonons.\cite{millis1986}
%%%%%%%%%%%%%%%%%%%%%%%%%%%%%%
	%%%%%%%%%%%%%%%%%%%%%%%%%%%%%%
	%%%%%%%%%%%%%%%%%%%%%%%%%%%%%%
	%%%%%%%%%%%%%%%%%%%%%%%%%%%%%%
	%%%%%%%%%%%%%%%%%%%%%%%%%%%%%%
	%%%%%%%%%%%%%%%%%%%%%%%%%%%%%%
	%%%%%%%%%%%%%%%%%%%%%%%%%%%%%%
	%%%%%%%%%%%%%%%%%%%%%%%%%%%%%%FIGURE2
\begin{figure}
\begin{centering}
\includegraphics[scale=0.41]{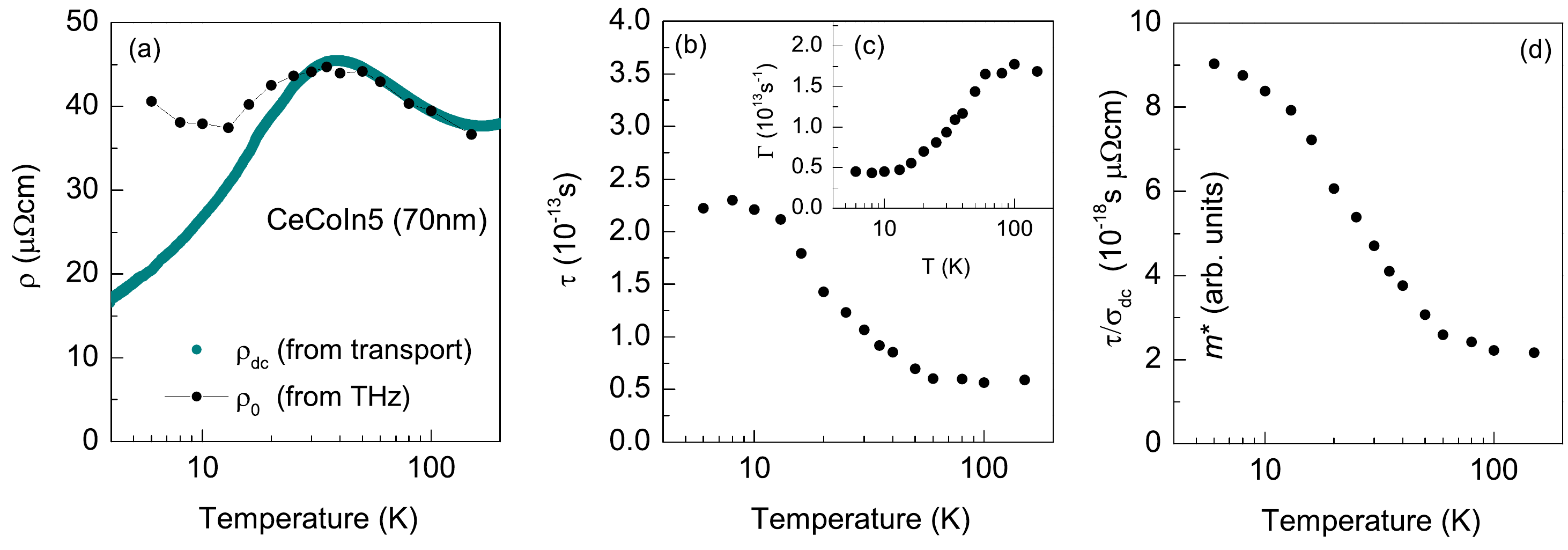}
\caption{\label{Drude} (Color online) Temperature dependence of (a) the dc resistivity obtained from optical and transport probes, (b) the scattering time $\tau$, (c) the scattering rate $\Gamma=1/\tau$, and (d) $\tau/\sigma_0$, which is a measure for the electron-mass enhancement. }
\end{centering}
\end{figure}
	%%%%%%%%%%%%%%%%%%%%%%%%%%%%%%
	%%%%%%%%%%%%%%%%%%%%%%%%%%%%%%
	%%%%%%%%%%%%%%%%%%%%%%%%%%%%%%
As further data analysis, we fitted the transmittance (see Fig. \ref{fits}(a) below) to well-established Fresnel equations for multiple reflections \cite{Dre02}, where the complex refractive index $n+ik$ is expressed in terms of the Drude conductivity 
\begin{equation}
\sigma_1(\omega)+i\sigma_2(\omega)=\sigma_0\left(\frac{1}{1+(\omega \tau)^2}+i\frac{\omega \tau}{1+(\omega \tau)^2}\right)\label{eq:Drude}  
\end{equation}
with angular frequency $\omega=2\pi \nu$ and temperature-dependent parameters dc conductivity $\sigma_0$ and scattering time $\tau=1/\Gamma$. Fig. \ref{Drude} (a) displays  $\rho_0=1/\sigma_0$, i.e.\ the dc resistivity from optical Drude analysis, and compares it to $\rho_{dc}$ obtained from the transport measurement. In the regime of incoherent Kondo scattering, i.e.\ between $T_{min}$ and $T_{max}$, the results of both the optical and transport measurements coincide. Between $T_{min}$ and the inflection point of the $\rho_{dc}(T)$ curve, $\tau$ and $\Gamma$, see Fig. \ref{Drude} (b) and inset (d), remain roughly constant. For lower temperatures, $\tau$ and $\Gamma$ tend to increase and decrease, respectively, signaling the emergence of the underlying heavy-fermion state. At around 25\,K, results from the optical and transport probes in Fig. \ref{Drude} (a) start to deviate. We observe a more rapid decrease of $\rho_{dc}$ compared to $\rho_0$ which even levels off at around 13\,K. Down to this temperature, $\tau$ and $\Gamma$ display a strong temperature dependence in the now well developed heavy-fermion state. At even lower temperatures, $\rho_0$ remains fairly constant and tends to a slight increase which is in clear contrast to the transport result. At the same time, the temperature dependence of $\tau$ and $\Gamma$ becomes less pronounced.
 In the Drude framework,  $\sigma_0$ is given by
\begin{equation}
\sigma_0=\frac{Ne^2\tau}{m}\label{eq:mass}
\end{equation}
where $N,e$ and $m$ are the electron density, charge, and mass. In this free electron picture, the temperature dependence of $\rho_0$ is usually determined by $\tau$ given that the number of carriers is constant. In heavy-fermion metals, however, $m$ is renormalized by electron-electron interactions and the effective mass $m^*$ becomes strongly temperature dependent. While $\rho_0$ remains fairly constant in the heavy-fermion regime, $\tau$ drastically increases. Even without knowledge of $N$ we can infer the temperature dependence of $m^*$ by plotting $\tau/\sigma_0=m^*/(Ne^2)$, see Fig. \ref{Drude}(c). Upon cooling, the mass enhancement already sets in well before the heavy-fermion state is fully developed, speeds up, and eventually levels off towards the lowest temperature. Assuming a constant value of $N$, this would translate to a mass enhancement of roughly four between the highest and lowest temperature of this study. This value is small when compared to those found for CeCoIn$_5$ from other techniques such as specific heat \cite{petrovic2001,kim2001} or quantum oscillations \cite{settai2001,mccollam2005}, but those probes usually reveal the effective mass only for very low temperatures, whereas our work indicates that the mass enhancement upon cooling is not complete yet at our lowest temperature of 6~K. Closer to our approach are optical studies at higher frequencies, in the infrared regime,\cite{singley2002,mena2005} which found a frequency-dependent effective mass enhancement which amounts to approximately 20 and 13, respectively, at their lowest frequencies and temperatures (30~cm$^{-1}$, 10~K and 40~cm$^{-1}$, 8~K, respectively). Since our optical measurements reach lower frequencies and temperatures, more pronounced mass enhancement is expected, which is not found in the data of Fig. \ref{Drude}(c). This can be explained since the simple Drude analysis performed here does not take into account any possible frequency dependence of scattering rate and effective mass. As we will show below, this description becomes more inaccurate for decreasing temperature. Here, a more detailed analysis based on data of the full optical response (real and imaginary parts) is desired.

	%%%%%%%%%%%%%%%%%%%%%%%%%%%%%%
	%%%%%%%%%%%%%%%%%%%%%%%%%%%%%%
%%%%%%%%%%%%%%%%%%%%%%%%%%%%%%
	%%%%%%%%%%%%%%%%%%%%%%%%%%%%%%
	%%%%%%%%%%%%%%%%%%%%%%%%%%%%%%
	%%%%%%%%%%%%%%%%%%%%%%%%%%%%%%
	%%%%%%%%%%%%%%%%%%%%%%%%%%%%%%

However, the trends that we find for $\rho_0, \tau, \Gamma$ and $m^*$ in the present work are in good agreement with previous results \cite{Sch13} on CeCoIn$_5$ thin films, where the analysis was less straightforward due to metallic buffer layers beneath the thin film. In \cite{Sch13} the discrepancy between optical and transport probes was explained by a rapid degradation of the film during the measurement. Here, this seems less likely as we could not observe any signs of degradation in the optical spectra. 

Fig. \ref{fits} (a) comprises the transmittance together with the Fresnel fits including the single-particle Drude formula, Eq. (\ref{eq:Drude}). At high temperatures, where no mismatch between optical and transport probes was visible, the fit captures the transmittance in the entire frequency range very well. Upon cooling down, the agreement between the simple theory and experimental data becomes less good: most severe deviations arise in the low and high frequency limits, where the actual transmittance drops below the theory expectation. Furthermore, a small phase shift arises at intermediate frequencies. By judging the fit fidelity, one can understand the discrepancy between $\rho_0$ and $\rho_{dc}$ as an increasing incapability of the single-particle Drude theory to reproduce the dynamical properties of CeCoIn$_5$. 
Indeed, in a number of correlated electron systems a frequency dependence of $\tau$ was found resulting from electron-electron interactions, and by Kramers-Kronig relations, one could also expect a frequency-dependent effective mass. This might also explain the failure of the Drude theory describing the many-body heavy-fermion state in CeCoIn$_5$ at sufficiently low temperatures.

In Fig. \ref{fits}(b) and (c) we qualitatively explain how such deviations from Drude behavior in the transmittance spectra can arise due to electronic correlations. Our reference for an interacting electron system is a Fermi liquid (FL) with optical properties well understood from the theoretical side \cite{gurzhi1959,maslov2012,Berthod2013} and recently studied experimentally.\cite{scheffler2013,schneider2014,stricker2014} For CeCoIn$_5$, non-FL behavior (with linear temperature dependence of dc resistivity compared to the FL quadratic behavior) was found experimentally for many properties and is expected for the THz response, but in lack of an appropriate non-FL prediction, we here refer to the FL case to demonstrate the overall behavior. FL theory predicts a quadratic frequency dependence of the scattering rate $\Gamma(\omega) = \Gamma(\omega=0) + b \omega^2$ (with the prefactor $b$ depending on material) compared to the frequency-independent $\Gamma$ within the Drude response. As a result, $\sigma_1(\omega)$ for FL notably deviates from the Drude case, with the characteristic feature being a higher $\sigma_1$ at higher frequency (\lq non-Drude trail\rq).\cite{Berthod2013} Such differences in $\sigma_1$ correspond to differences in the transmittance, as is shown in Fig. \ref{fits}(c): if one compares the transmittance spectra of an interacting electron system (our CeCoIn$_5$ data at low temperature and the FL in the schematic figure) with the spectrum for a Drude metal, one finds that the maxima of the FP oscillations can be modeled properly in a limited (intermediate) frequency range, while the maxima in the Drude case surpass those of the interacting case for both lower and higher frequency. Therefore we interpret the insufficient Drude fits of our low-temperature transmittance data as evidence for electronic correlations in the THz response of CeCoIn$_5$. Whether these can be described as FL optics or whether, as expected, non-FL features govern the THz properties remains to be seen from further studies that should address the phase shift in the THz response in addition to the transmittance.

	%%%%%%%%%%%%%%%%%%%%%%%%%%%%%%
	%%%%%%%%%%%%%%%%%%%%%%%%%%%%%%
	%%%%%%%%%%%%%%%%%%%%%%%%%%%%%%
	%%%%%%%%%%%%%%%%%%%%%%%%%%%%%%
	%%%%%%%%%%%%%%%%%%%%%%%%%%%%%%

\section{Summary}
\label{s}
In summary, we discussed the transmittance of THz radiation through a high-quality thin film of CeCoIn$_5$ measured by quasi-optical spectroscopy and compared it to transport measurements of the dc  resistivity $\rho_{dc}$. We found a perfect agreement of the dc resistivity $\rho_0=1/\sigma_0$ obtained from Drude optics and $\rho_{dc}$ in the regime of incoherent Kondo scattering. At lower temperatures, the scattering time $\tau$ and effective mass $m^*$ acquire a strong temperature dependence and the agreement between  $\rho_{dc}$ and $\rho_0$ becomes worse. We attribute this to an increasing incapability of the simple single-particle picture, i.e.\ the Drude theory, in favor of a more advanced description that accounts for the electronic correlations associated with the low-temperature heavy-fermion state. With the recent improvements in growing high-quality thin films, optical experiments at THz and GHz frequencies become feasible and we hope that our results motivate further investigations illuminating the unconventional charge carrier dynamics in CeCoIn$_5$.
	%%%%%%%%%%%%%%%%%%%%%%%%%%%%%%
	%%%%%%%%%%%%%%%%%%%%%%%%%%%%%%
	%%%%%%%%%%%%%%%%%%%%%%%%%%%%%%
	%%%%%%%%%%%%%%%%%%%%%%%%%%%%%%
	%%%%%%%%%%%%%%%%%%%%%%%%%%%%%%
	%%%%%%%%%%%%%%%%%%%%%%%%%%%%%%

\section{Acknowledgements}
\label{a}
This study was supported by the DFG. The work in Japan was supported by KAKENHI from JSPS. U.S.P. acknowledges financial support from the Studienstiftung des deutschen Volkes. 
	%%%%%%%%%%%%%%%%%%%%%%%%%%%%%%
	%%%%%%%%%%%%%%%%%%%%%%%%%%%%%%
	%%%%%%%%%%%%%%%%%%%%%%%%%%%%%%
	%%%%%%%%%%%%%%%%%%%%%%%%%%%%%%
	%%%%%%%%%%%%%%%%%%%%%%%%%%%%%%
	%%%%%%%%%%%%%%%%%%%%%%%%%%%%%%

\end{document}